\documentclass[9pt,conference]{IEEEtran}

\usepackage{waspaa25} 

\usepackage{bm} 
\usepackage{tcolorbox} 
\definecolor{lightgray}{rgb}{0.9, 0.9, 0.9} 
\usepackage{xspace}

\newcommand{\modelname}{\text{\textbf{LLM2Fx}}\xspace}

\usepackage{listings}

\lstdefinestyle{json}{
    basicstyle=\ttfamily\small,
    numbers=none,
    breaklines=true,
    morestring=[b]",  
    stringstyle=\color{green!50!black},
    showstringspaces=false,
    tabsize=2,
    frame=single,
    backgroundcolor=\color{gray!5},
    moredelim=**[is][\color{blue}]{@}{@},  
    moredelim=**[is][\color{red}]{|}{|},   
    captionpos=b
}

\lstdefinestyle{docstring}{
    basicstyle=\ttfamily\small,
    breaklines=true,
    frame=single,
    backgroundcolor=\color{gray!5},
    numbers=none,
    columns=flexible,
    keepspaces=true
}

\title{Can Large Language Models Predict Audio Effects Parameters\\from Natural Language?}

\name{Seungheon Doh$^{1,2,*}$~\thanks{*Work done while an intern at Sony AI},
      Junghyun Koo$^{2}$,
      Marco A. Martínez-Ramírez$^{2}$,
      Wei-Hsiang Liao$^{2}$,
      Juhan Nam$^{1}$
      Yuki Mitsufuji$^{2,3}$
}
\address{$^{1}$KAIST, South Korea \;
$^{2}$Sony AI, Japan \;
$^{3}$Sony Group Corporation, Japan
}

\begin{document}

\maketitle

\begin{abstract}
In music production, manipulating audio effects~(Fx) parameters through natural language has the potential to reduce technical barriers for non-experts. We present \modelname, a framework leveraging Large Language Models (LLMs) to predict Fx parameters directly from textual descriptions without requiring task-specific training or fine-tuning. Our approach address the text-to-effect parameter prediction (Text2Fx) task by mapping natural language descriptions to the corresponding Fx parameters for equalization and reverberation. We demonstrate that LLMs can generate Fx parameters in a zero-shot manner that elucidates the relationship between timbre semantics and audio effects in music production. To enhance performance, we introduce three types of in-context examples: audio Digital Signal Processing (DSP) features, DSP function code, and few-shot examples. Our results demonstrate that LLM-based Fx parameter generation outperforms previous optimization approaches, offering competitive performance in translating natural language descriptions to appropriate Fx settings. Furthermore, LLMs can serve as text-driven interfaces for audio production, paving the way for more intuitive and accessible music production tools.~\footnote{Our demo is available at \url{https://seungheondoh.github.io/llm2fx-demo/}}
\end{abstract}

\section{Introduction}
Semantic music production represents an emerging paradigm that leverages natural language descriptions to control audio effects~(Fx)~\cite{moffat2022semantic}. This approach aims to reduce the technical barriers associated with traditional music production interfaces, making the creative process more accessible to novices and non-experts. The text-to-Fx parameter prediction (Text2Fx) task stands at the forefront of this innovation, enabling users to articulate desired sonic transformations in everyday language (e.g., ``Make the guitar sound warmer'' or ``Add a bright, spacious reverb'') rather than navigating complex technical parameters or searching for reference audio samples.

Prior research in semantic audio production has made significant strides in connecting natural language with audio processing parameters. Sabin et al.~\cite{sabin2011weighted} pioneered the mapping of text descriptors to Equalizer~(EQ) and Reverberation~(Reverb) parameters. The SAFE project~\cite{stables2016semantic} expanded this work by developing a comprehensive dataset and ontological framework linking language to audio transformations. The SocialFX series~\cite{cartwright2013social, seetharaman2014reverbalize, zheng2016socialfx} contributed valuable insights through crowdsourced natural language descriptions for EQ, Reverb, and Compression. The Mix Evaluation Dataset~\cite{de2017mix} collected mixes including parameter settings, preference ratings, and free-form descriptions.

For the Text2Fx task, Venkatesh et al.~\cite{venkatesh2022word} proposed an EQ parameter estimation approach using effects-specific word embedding spaces. Balasubramaniam et al.~\cite{balasubramaniam2023word} introduced text-driven audio manipulation, but their black-box approach limited user control and understanding while being constrained by a closed vocabulary. Chu et al.~\cite{chu2025text2fx} addresses these limitations by leveraging CLAP~\cite{elizalde2023clap} embeddings and differentiable DSP modules~\cite{steinmetz2022style, steinmetz2021filtered} to enable open-vocabulary control of Fx. Their method performs single-instance optimization in the embedding space without requiring model retraining, offering a flexible approach to sound transformation.

Beyond regression or optimization approaches, the Text2Fx task can be addressed through generative methods. One promising approach is to leverage LLMs for direct Fx parameter prediction, which can potentially offer more flexible solutions compared to traditional techniques. Recent advancements in LLMs have demonstrated remarkable zero-shot capabilities across various domains. LLMs like GPT-4~\cite{achiam2023gpt}, Gemini~\cite{team2024gemini}, and LLaMA series~\cite{grattafiori2024llama} have shown the ability to perform complex tasks, leveraging their extensive pre-training on diverse text corpora. This zero-shot generalization has been particularly effective in domains requiring specialized knowledge and reasoning~\cite{wei2022emergent}. In the music and audio domain, while LLMs have been successfully applied to tasks such as understanding~\cite{doh2023lp, gardner2023llark, tang2023salmonn, liu2024music, zhao2024openmu}, retrieval~\cite{doh2024enriching, manco2024augment, doh2024music, wu2025clamp, doh2025talkplay}, and generation~\cite{yuan2024chatmusician} tasks, they have not yet been extensively explored for music post-production tasks.


\begin{figure}[!t]
\centering
\includegraphics[width=0.90\linewidth]{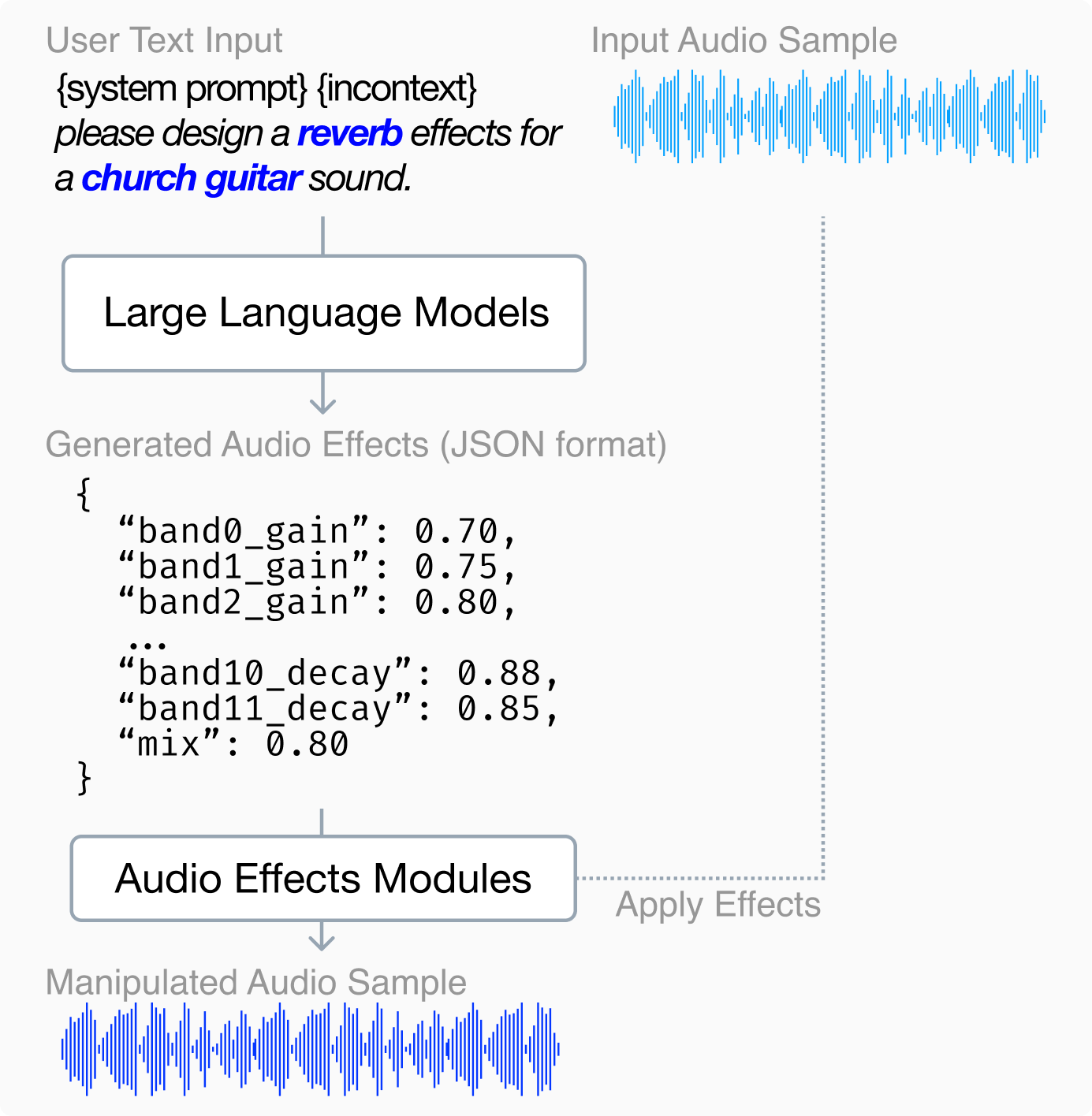}
\vspace{-2mm}
\caption{Illustration of the $\modelname$ Framework. The user inputs natural language descriptions for music production tasks, and the large language model predicts appropriate Fx parameters based on 1) the given input, 2) system prompt, and 3) in-context information. The predicted parameters are parsed and then applied to Audio Effect Modules to transform the sound.}
\vspace{-1mm}
\end{figure}

In this work, we investigate the capabilities of LLMs as semantic audio experts for the Text2Fx task. Our contributions are threefold: 1) We demonstrate that LLMs can generate plausible Fx parameters settings directly from natural language descriptions without task-specific training; 2) We introduce novel in-context inference strategies specifically designed for audio processing tasks that significantly enhance parameter generation performance; 3) We develop a comprehensive quantitative evaluation pipeline for the Text2Fx task, providing analysis of existing datasets and multiple audio embedding extractors. Our approach offers a promising direction for creating more intuitive and accessible music production interfaces that leverages the semantic understanding capabilities of LLMs.

\section{LLM2Fx: LLM-based Audio~Effects~Parameter~Prediction}

\subsection{Task Formation}
We formalize the Text2Fx task as a conditioned prediction problem. Given a natural language description, the LLM predicts appropriate audio effect parameters $p = f_{\text{LLM}}(d, \mathcal{P}, \mathcal{C})$, where $d$ represents natural language descriptions of desired timbre and audio effect transformations (e.g., ``make the sound warmer''), $p$ denotes the predicted audio effect parameter values in structured JSON format, $\mathcal{P}$ is the system prompt that guides the model's behavior, and $\mathcal{C}$ represents in-context information.

The model generates Fx parameters that appropriately realize the requested timbral characteristics for EQ and Reverb. We employ pre-trained LLMs as our mapping function $f_\text{LLM}$ based on their demonstrated zero-shot reasoning capabilities across specialized knowledge domains. We hypothesize that these models have acquired implicit knowledge of audio processing concepts through their pre-training on diverse text corpora, enabling them to effectively bridge the semantic gap between natural language descriptions and technical audio parameters without requiring task-specific training or fine-tuning.

\subsection{System Prompting}
To effectively leverage LLMs for Fx parameter generation, we design a specialized system prompt that frames the task and provides necessary constraints. Our system prompt serves three key functions: \textit{1) Role Definition}: establishes the LLM as an expert audio engineer with specialized knowledge in sound design and audio processing; \textit{2) Task Instruction}: clearly defines the objective of translating natural language description, such as  semantic word, instrument type, Fx type, into specific audio effect parameters; and \textit{3) Response Format}: enforces structured output in JSON format to ensure parameter predictions are machine-readable. This prompt design ensures that the LLM generates structured parameter predictions while leveraging its implicit knowledge of audio processing concepts.

\subsection{In-Context Information}
To enhance the performance of LLMs on the Text2Fx task, we introduce three types of in-context information that provide domain knowledge for audio processing tasks.

\vspace{2mm}
\noindent \textbf{Few-Shot Examples Context:} We utilize the in-context learning capabilities~\cite{brown2020language} of LLMs by providing few-shot examples that demonstrate the mapping between natural language descriptions and corresponding audio effect parameters. Formally, we can express this as: $f_{\text{LLM}}(d, \mathcal{P}, \mathcal{C}_{\text{few-shot}}) = p$ where $\mathcal{C}_{\text{few-shot}} = \{(d_1, p_1), (d_2, p_2), \ldots, (d_k, p_k)\}$ represents $k$ examples of semantic descriptions $d_i$ paired with their corresponding parameter settings $p_i$. Each example consists of a semantic description (e.g., ``warm piano sound'') paired with appropriate parameter settings in JSON format. These examples serve as implicit guidance for the model, helping it understand the relationship between semantic descriptors and technical parameters without requiring explicit fine-tuning. For few-shot examples, we utilize five manually annotated human examples ($k=5$) for each audio effect type, ensuring high-quality reference points for the LLMs.

\vspace{2mm}
\noindent \textbf{DSP Function Context:} We leverage the code comprehension abilities~\cite{achiam2023gpt, team2024gemini, roziere2023code} of LLMs by including DSP code function at inference time. Using the DASP Python library~\footnote{https://github.com/csteinmetz1/dasp-pytorch}, we include code snippets that explicitly show the mathematical relationships between parameters and their effect on the audio signal. By exposing the model to these implementations, we enable it to reason about the underlying signal transformations associated with different parameter settings, leading to more technically sound predictions.

\vspace{2mm}
\noindent \textbf{DSP Feature Context:} To enhance parameter prediction specifically conditioned on input audio, we provide statistical information of audio DSP features as contextual examples. This approach explores whether adding these parameters as context might improve how LLMs predict audio effects, as blind prediction of audio effects is highly unusual in audio engineering. Given that LLMs might have been trained on extensive literature regarding DSP low-level descriptors and high-level semantic words, we hypothesize that this exposure could help LLMs explore the relationship between these topics. We include representative audio features extracted from the input signal as context for LLM inference, such as:

\begin{itemize}
    \item Sample rate: Number of samples per second
    \item RMS energy: $\sqrt{\frac{1}{N}\sum_{n=0}^{N-1}x^2[n]}$, where $x[n]$ is the audio signal
    \item Crest factor: $\frac{\max(|x[n]|)}{RMS}$, ratio of peak amplitude to RMS
    \item Dynamic spread: $\sqrt{\frac{1}{N}\sum_{n=0}^{N-1}(|x[n]| - \mu)^2}$, where $\mu$ is the mean amplitude
    \item Spectral centroid: $\frac{\sum_{k=0}^{N/2}f[k]|X[k]|}{\sum_{k=0}^{N/2}|X[k]|}$, where $X[k]$ is the FFT of $x[n]$ and $f[k]$ is the frequency at bin $k$
    \item Spectral flatness: $\frac{\sqrt[N]{\prod_{k=0}^{N/2}|X[k]|}}{\frac{1}{N}\sum_{k=0}^{N/2}|X[k]|}$, ratio of geometric to arithmetic mean of spectrum
    \item Spectral bandwidth: $\sqrt{\frac{\sum_{k=0}^{N/2}(f[k]-SC)^2|X[k]|}{\sum_{k=0}^{N/2}|X[k]|}}$, spread of frequencies around the centroid
    \item RT60: time for sound to decay by 60dB, estimated reverberation time of an impulse response using the Schroeder method~\cite{schroeder1965new}.
\end{itemize}

\noindent These DSP features provide a numerical representation of the audio signal's characteristics that can be included as context for the LLM. Our pre-trained LLMs can only process natural language. However, we expect that providing DSP features expressed in language will help the model generate audio-specific effect parameters. These parameters will be conditioned on both the semantic word and the input audio content. This approach creates a bridge between audio and language modalities without needing explicit multimodal training.

\vspace{2mm}
\section{Experiments}

We evaluated the \modelname framework on two Fx: EQ and Reverb, explicitly using the 6-band Parametric EQ (18 parameters) and the Noise-Shaped Reverberation (25 parameters) implementations from the DASP library.

\subsection{Evaluation Dataset}
An ideal dataset for semantic music–production research should provide four aligned modalities: a dry (unprocessed) audio signal, Fx parameters, the resulting wet (processed) signal, and the natural-language description of the effected audio. Two public datasets currently satisfy these requirements—\emph{SocialEQ}~\cite{cartwright2013social} and \emph{SocialReverb}~\cite{seetharaman2014reverbalize}. \emph{SocialEQ} records one descriptor per example (multi-class), whereas \emph{SocialReverb} may attach several descriptors to the same example (multi-label). These two datasets were merged into the \emph{SocialFX} dataset~\cite{zheng2016socialfx} and are accessible as open-source resources\footnote{\url{http://bit.ly/1WmTP6v}}. Based on \emph{SocialFX} dataset, we reproduced all audio examples using the original parameters with the \emph{Audealize} JavaScript toolkit~\cite{seetharaman2016audealize}\footnote{\url{https://github.com/interactiveaudiolab/audealize_api}}. Following the methodology and using samples from \cite{seetharaman2016audealize}, we applied each parameter set to dry samples of three instruments—\textit{guitar} (10 seconds), \textit{drums} (15 seconds), and \textit{piano} (20 seconds)—to generate wet versions that differ only in Fx, preserving consistent musical content within each instrument class.

It is important to note that semantic annotations (such as ``\textit{warm}'' or ``\textit{bright}'') represent highly subjective concepts. The relationship between audio and semantic words is inherently many-to-many: a single descriptor can correspond to a variety of audio characteristics, and a given audio sample may be described in multiple terms. To support reliable quantitative evaluation, we ensured that each semantic word contains a sufficient number of Fx samples. This was achieved through a three-step preprocessing pipeline: 1) word merging, 2) term frequency filtering, and 3) classification score filtering. Initially, we had 110 words for EQ and 71 words for Reverb. After the three-step preprocessing, this was reduced to 7 and 19 words, respectively.

\vspace{2mm}
\noindent \textbf{Word Merging:} We used GPT-4o~\cite{achiam2023gpt} to create initial clusters of semantically similar words from the entire vocabulary. Using this LLM-based pseudo-clustering as a starting point, we established word merging rules through human validation to confirm semantic similarity (i.e, $\textit{warm, heat, toasty}\xrightarrow{}\textit{warm}$). For each cluster, we selected the word with the highest term frequency as the representative word. This process significantly reduced the vocabulary size to 26 words for EQ and 19 words for Reverb.

\vspace{2mm}
\noindent \textbf{Term Frequency Filtering:} We filtered out words with insufficient audio effect samples to ensure reliable evaluation. Specifically, we used thresholds of 20 samples for EQ and 100 samples for Reverb, removing any words that fell below these minimum requirements. As a result, we obtained 9 words for EQ (\textit{warm, soft, harsh, calm, loud, bright, heavy, happy, cool}) and 19 words for reverb (\textit{echo, distant, spacious, muffled, warm, church, loud, deep, soft, distorted, sad, hollow, sharp, low, tinny, underwater}).

\begin{table}[!t]
\centering
\caption{Statistics of the SocialFX dataset before preprocessing (Original) and after preprocessing (Eval) for EQ and Reverb.}
\label{tab:table1}
\sisetup{
    reset-text-series = false,
    text-series-to-math = true,
    mode=text,
    tight-spacing=true,
    round-mode=places,
    round-precision=2,
    table-format=2.2,
    table-number-alignment=center
}
\begin{tabular}{lcclcc}
\toprule
\textbf{} & \multicolumn{2}{c}{\textbf{EQ}} &  & \multicolumn{2}{c}{\textbf{Reverb}} \\ \cmidrule{2-3} \cmidrule{5-6}
Statistics & Original~\cite{cartwright2013social} & Eval &  & Original~\cite{seetharaman2014reverbalize} & Eval \\ \midrule
Parameter Sets & 1595 & 273 &  & 6772 & 3833 \\
Vocabulary Size & 765 & 7 &  & 3535 & 19 \\
Avg. Effect per Word & 2.1 & 39.0 &  & 5.3 & 291.7 \\
Avg. Word per Effect & 1.0 & 1.0 &  & 2.8 & 1.4 \\ \bottomrule
\end{tabular}
\end{table}

\vspace{2mm}
\noindent \textbf{Classification Score Filtering:} To ensure strong associations between semantic words and manipulated audio samples, we implemented a classification-based filtering approach. We extracted embeddings from manipulated audio samples using available pre-trained audio embedding extractors, then trained a linear probing layer to predict semantic words. For this linear probing task, we utilized four different audio embedding extractors: \text{VGGish}~\cite{hershey2017cnn}, \text{CLAP}~\cite{wu2023large}, \text{AFx-Rep}~\cite{steinmetz2024st}, and \text{FXenc}~\cite{koo2023music}. We compared the F1 scores of these classifiers with those of a gaussian random embedding baseline, removing any labels whose performance fell below the random baseline. We compared the average F1 scores of these classifiers against a gaussian random embedding baseline, removing semantic words whose performance fell below this threshold. When classifiers underperform the random baseline, it indicates the audio embeddings lack meaningful information related to that semantic word. This filtering process eliminated abstract terms like ``happy'' and ``cool'' from the EQ vocabulary. The final result was a vocabulary of 7 words for EQ and 19 words for Reverb.

Table~\ref{tab:table1} shows the results before and after preprocessing. We named our preprocessed dataset \emph{SocialFX-Eval}. While we end up with fewer audio effect and natural language pairs, we obtain a more sufficient number of samples per word, enabling more robust evaluation.

\begin{table}[!t]
\centering
\caption{MMD scores across different audio features. U.B represents upper bound (lower is better), L.B represents lower bound (upper is better), and $\Delta$ shows the difference between U.B and L.B (lower is better). Avg. $\Delta$ stands for average value of EQ $\Delta$ and Reverb $\Delta$}
\resizebox{1.025\linewidth}{!}{%
\label{tab:table2}
\sisetup{
    reset-text-series = false,
    text-series-to-math = true,
    mode=text,
    tight-spacing=true,
    round-mode=places,
    round-precision=2,
    table-format=2.2,
    table-number-alignment=center
}
\begin{tabular}{lccclcccc}
\toprule
 & \multicolumn{3}{c}{\textbf{EQ}} & \textbf{} & \multicolumn{3}{c}{\textbf{Reverb}} &  \\ \cmidrule{2-4} \cmidrule{6-8}
Models & U.B & L.B & $\Delta$  &  & U.B & L.B & $\Delta$ & Avg.$\Delta$ \\ \midrule
DSP feat. & 0.35 & 0.23 & 0.12 &  & 0.14 & 0.17 & -0.03 & 0.04 \\
CLAP~\cite{wu2023large} & 0.07 & 0.53 & \textbf{-0.46} &  & 0.07 & 0.40 & -0.34 & -0.40 \\
VGGish~\cite{hershey2017cnn} & 0.16 & 0.62 & \textbf{-0.46} &  & 0.09 & 0.52 & -0.43 & -0.44 \\
AFx-Rep~\cite{steinmetz2024st} & 0.18 & 0.57 & -0.40 &  & 0.09 & 0.41 & -0.32 & -0.36 \\
FXenc~\cite{koo2023music} & 0.09 & 0.53 & -0.44 &  & 0.05 & 0.75 & \textbf{-0.70} & \textbf{-0.57} \\ \bottomrule
\end{tabular}
}
\end{table}

\begin{table*}[!t]
\centering
\caption{MMD scores comparing Text2FX performance across different models and instruments. For MMD scoring, we use FXenc~\cite{koo2023music} for backbone embedding model. Lower scores indicate better alignment between predicted and ground truth distributions. (lower is better)}
\label{tab:results}
\sisetup{
    reset-text-series = false,
    text-series-to-math = true,
    mode=text,
    tight-spacing=true,
    round-mode=places,
    round-precision=2,
    table-format=2.2,
    table-number-alignment=center
}
\begin{tabular}{lllcccclcccc}
\toprule
 &  &  & \multicolumn{4}{c}{\textbf{EQ}} &  & \multicolumn{4}{c}{\textbf{Reverb}} \\ \cmidrule{4-7} \cmidrule{9-12} 
Models & Methods &  & Drums  & Guitar & Piano & Avg. &  & Drums & Guitar & Piano & Avg. \\ \midrule
Lowerbound & Random Parameters &  & 0.45 & 0.49 & 0.65 & 0.53 &  & 0.72 & 0.69 & 0.84 & 0.75 \\
Chu et al.~\cite{chu2025text2fx} & Optimization with &  & 0.49 & 0.42 & 0.75 & 0.55 &  & 0.41 & 0.42 & 0.53 & 0.45 \\ \midrule
GPT-4o~\cite{achiam2023gpt} & Zeroshot Generation &  & \textbf{0.22} & \textbf{0.15} & \textbf{0.28} & \textbf{0.22} &  & 0.66 & 0.64 & 0.79 & 0.70 \\
Llama3.2-1B-Instruct~\cite{grattafiori2024llama} & Zeroshot Generation &  & 0.56 & 0.62 & 0.73 & 0.64 &  & 0.63 & 0.61 & 0.75 & 0.66 \\
Llama3.2-3B-Instruct~\cite{grattafiori2024llama} & Zeroshot Generation &  & 0.24 & 0.20 & 0.37 & 0.27 &  & 0.60 & 0.53 & 0.72 & 0.61 \\
Llama3.1-8B-Instruct~\cite{grattafiori2024llama} & Zeroshot Generation &  & 0.25 & 0.20 & 0.31 & 0.25 &  & 0.65 & 0.65 & 0.74 & 0.68 \\
Llama3.3-70B-Instruct~\cite{grattafiori2024llama} & Zeroshot Generation &  & 0.25 & 0.19 & 0.29 & 0.24 &  & 0.37 & 0.47 & 0.72 & 0.52 \\
Mistral-7B-Instruct-v03~\cite{jiang2023mistral} & Zeroshot Generation &  & 0.29 & 0.26 & 0.35 & 0.30 &  & \textbf{0.35} & \textbf{0.41} & \textbf{0.59} & \textbf{0.45} \\ \midrule
\hspace{1mm} + DSP Audio Features & Zeroshot Generation &  & 0.29 & 0.22 & 0.30 & 0.27 &  & 0.49 & 0.52 & 0.63 & 0.55 \\
\hspace{1mm} + DSP Python Code & Zeroshot Generation &  & 0.22 & 0.20 & 0.28 & 0.23 &  & 0.71 & 0.75 & 0.85 & 0.77 \\
\hspace{1mm} + Fewshot Examples & Fewshot Generation &  & \textbf{0.21} & 0.23 & 0.27 & 0.23 &  & \textbf{0.22} & 0.30 & 0.32 & 0.28 \\
\hspace{1mm} + All Context Info. & Fewshot Generation &  & 0.22 & \textbf{0.21} & \textbf{0.23} & \textbf{0.22} &  & 0.24 & \textbf{0.26} & \textbf{0.30} & \textbf{0.27} \\

\midrule
{\color[HTML]{656565} Upperbound} & {\color[HTML]{656565} Ground Truth} & {\color[HTML]{656565} } & {\color[HTML]{656565} 0.09} & {\color[HTML]{656565} 0.07} & {\color[HTML]{656565} 0.12} & {\color[HTML]{656565} 0.09} & {\color[HTML]{656565} } & {\color[HTML]{656565} 0.05} & {\color[HTML]{656565} 0.05} & {\color[HTML]{656565} 0.04} & {\color[HTML]{656565} 0.05}   \\ \bottomrule
\end{tabular}
\end{table*}

\subsection{Evaluation Protocol}
Our evaluation process involves an LLM predicting Fx parameters from natural language descriptions, which are then applied to input audio samples via DASP Fx modules. Although our ground truth Fx modules were JavaScript implementations, we use DASP modules as prediction targets to enable fair comparison with prior work~\cite{chu2025text2fx}. We conduct 50 trials to generate different parameter sets for each semantic word. For evaluation metric, we employ Maximum Mean Discrepancy (MMD)~\cite{gretton2012kernel}, which measures the distance between two distributions: (1) audio features from samples processed with LLM-generated parameters and (2) audio features from samples processed with ground truth parameters for the same semantic word. MMD is defined as:
$\text{MMD}^2(P,Q) = \mathbb{E}_{x,x' \sim P}[k(x,x')] + \mathbb{E}_{y,y' \sim Q}[k(y,y')] - 2\mathbb{E}_{x \sim P, y \sim Q}[k(x,y)]$
where $k$ is a gaussian kernel function that quantifies the distance between distributions in a high-dimensional feature space. MMD scores enable reliable evaluation with small size of reference data~\cite{jayasumana2024rethinking, chung2025kad}. Lower MMD scores indicate better alignment between predicted and ground truth distributions, suggesting the LLM-generated parameters successfully capture the acoustic characteristics implied by the semantic word.

To select appropriate audio embedding extractors for our evaluation, we established upper- and lower-bounds to measure the distribution modeling capability of each method. The upper bound (U.B.) is the MMD between two sets of audio samples synthesized from ground-truth parameters for the same semantic word, obtained by splitting the parameter set in half. The lower bound (L.B.) is the MMD between audio samples from ground-truth parameters and those from randomly sampled parameters. As shown in Table~\ref{tab:table2}, we evaluated five different feature extractors: DSP features, CLAP, VGGish, AFx-Rep, and FXenc. We conducted sampling with 5 random seeds for both upper and lower bounds and reported the average values. The difference between upper and lower bounds ($\Delta$) served as our primary selection criterion. This demonstrates that the distribution of semantic clusters exhibits significant differences compared to random parameter distributions. For EQ, CLAP and VGGish demonstrated the largest difference (-0.46), while FXenc showed the strongest discrimination for reverb (-0.70). When considering the average difference across both Fx types, FXenc achieved the best overall performance (-0.57). Based on these results, we selected FXenc for evaluation in our subsequent experiments.

\subsection{Baseline Models}
We compare \modelname framework against an optimization-based approach to Fx parameter prediction. Chu et al.~\cite{chu2025text2fx} leverages CLAP~\cite{elizalde2023clap} embedding space and gradient-based optimization to find Fx parameters that maximize similarity between processed audio and target descriptions. We compare against various open-source LLMs, including LLaMA3 series
~\cite{grattafiori2024llama} (1B, 3B, 8B, 70B parameter versions) and Mistral~\cite{jiang2023mistral} (7B parameters), as well as the closed-source GPT-4o accessed through API calls. This comprehensive comparison allows us to assess how different model architectures and parameter scales affect audio effect parameter prediction quality, while maintaining the same system prompt structure and evaluation protocol across all LLM-based approaches to ensure fair comparison.

\section{Results}

\subsection{Can LLMs generate appropriate audio effects parameters from natural language?}
Results in Table~\ref{tab:results} demonstrate that LLMs can effectively generate appropriate Fx parameters from natural language descriptions. For EQ, GPT-4o achieves an average MMD score of 0.22, significantly outperforming both random parameter selection (0.53). However, for reverb, while Mistral-7B-Instruct substantially improves over random parameters (0.75), it only matches the optimization approach~\cite{chu2025text2fx} (both 0.45). These results indicate LLMs have acquired implicit knowledge about Fx during pre-training, with a notable disparity between Fx types.

\subsection{Does model size affect Text2Fx prediction performance?}
The impact of model scale on parameter generation quality is evident when comparing performance across different model sizes. For EQ, we observe a dramatic improvement from Llama3.2-1B-Instruct (0.64) to Llama3.2-3B-Instruct (0.27), with further gains as model size increases. Llama3.3-70B-Instruct achieves an average MMD score of 0.24, approaching GPT-4o's performance (0.22). For reverb, larger models generally perform better, with Llama3.3-70B-Instruct (0.52) substantially outperforming smaller Llama variants. Interestingly, Mistral-7B-Instruct achieves the best reverb performance (0.45) despite its smaller parameter count, suggesting that model architecture and training dataset also play important roles. The relationship between model scale and parameter quality indicates that audio signal processing knowledge is encoded more comprehensively in larger models.

\subsection{Does in-context information influence the Fx parameters generation?}
The bottom section of Table~\ref{tab:results} demonstrates the substantial impact of in-context information on parameter generation quality. Using Mistral-7B-Instruct as the base model, we observe adding different types of contextual information provide varying benefits. Adding DSP features of input audio improves EQ performance (from 0.30 to 0.27) but degrades reverb performance (from 0.45 to 0.55). Providing DSP Python code yields modest improvements for EQ parameters (0.30 to 0.23) but significantly degrades reverb performance (0.45 to 0.77). This disparity can be attributed to the fundamental differences between these Fx: while EQ modules follow relatively standardized implementations across audio systems~\cite{valimaki2016all}, reverberation algorithms exhibit substantial variation in design and implementation details~\cite{valimaki2012fifty}. These results suggest that simpler Fx, like EQ, where implementation variations have relatively less impact, are better suited to utilizing the pre-trained knowledge of current LLMs. Conducting experiments with various reverb implementations might further validate the observed performance disparity. 

However, when DSP Python code is combined with input audio features and few-shot examples, we observe significant performance improvements for both Fx. The complementary nature of these context types allows the model to understand both the algorithmic implementation and real-world parameter mappings, resulting in the best overall performance (0.22 for EQ and 0.27 for reverb). This suggests that while code alone may be insufficient, its combination with few-shot examples creates a powerful framework for Fx parameter prediction.

\section{Conclusion}
In this work, we introduced \modelname, demonstrating that LLMs can effectively translate natural language descriptions into Fx parameters without requiring task-specific training. Our experiments reveal that in-context information—particularly the synergistic combination of DSP features, function code, and few-shot examples—substantially enhances prediction performance by providing both algorithmic understanding and practical parameter mappings. Despite these findings, important limitations remain. while we performed data cleaning, \textit{semantic words-Fx parameters} mapping dataset is still requires further refinement to establish more robust connections. Additionally, our approach currently processes only a single Fx rather than a chain of Fx and works with single-track instrument sources rather than multitrack music. Furthermore, we used only one sample per instrument class, so further analysis is needed to fully assess the LLMs' ability to generalize Fx parameter predictions.

\section{Acknowledgment}
\label{sec:ack}
We thank Annie Chu for valuable discussions and insights regarding the SocialFX dataset access and evaluation protocols.



\bibliographystyle{IEEEtran}
\bibliography{reference}

\newpage
\onecolumn

\appendix
This appendix illustrates the standardized input-output protocol implemented in the \modelname framework. The framework utilizes natural language queries for audio effect parameter generation. A representative example query is shown below:
$$
\texttt{"Please design a \{fx\_type\} audio effect for a \{timbre\_word\} \{instrument\} sound."}
$$
The system then generates appropriate parameters based on the description. We show a reverb example to provide an intuitive understanding of how our framework processes natural language descriptions into audio effect parameters.

\subsection{Input example of LLM inference}

\begin{lstlisting}[style=docstring]
You are an expert audio engineer and music producer specializing in sound design and audio processing. Your task is to translate descriptive timbre words into specific audio effects parameters that will achieve the desired sound character. You have deep knowledge of equalizers and understand how they shape timbre. You MUST respond with ONLY a valid JSON object.

# Instruction Format
Given a reverb description word or phrase and an instrument type, generate appropriate parameters for a frequency-dependent reverb that will achieve the requested spatial character.
For 44100 sample rate audio, Consider the typical reverb needs of the specified instrument when designing the reverb characteristics.

# Input Format
The input will consist of:
1. A reverb description such as:
   - Single words: "hall", "room", "plate", "cathedral", "chamber", "spring", "ambient"
   - Combined descriptions: "warm hall", "bright room", "dark chamber", "short but dense"
   - Spatial descriptions: "distant", "close", "intimate", "huge", "airy", "tight"
2. An instrument type such as:
   - "drums", "guitar", "piano", "vocals", "strings", "brass"

# Output Format
Respond with a JSON object containing precise numerical parameters for the reverb. All values should be in float format for efficiency. The output will include:
- The reverb parameters optimized for the requested spatial character and instrument. All values should be floating point numbers with 2 decimal places of precision.
Format:
{
    "reverb": {
        "band0_gain": float,
        "band1_gain": float,
        "band2_gain": float,
        "band3_gain": float,
        "band4_gain": float,
        ...
        [THE REST OF THE PARAMETERS ARE OMITTED]
    }
}

# Signal processing function

import numpy as np
import math
import scipy.signal
from scipy.fft import rfft, irfft
from functools import partial

def noise_shaped_reverberation(
    x,  # Shape: (channels, seq_len)
    sample_rate,
    band0_gain,
    ...
    band11_decay,
    mix
):
[THE REST OF THE PYTHON CODE ARE OMITTED]

# Input audio feature

{
    "sample_rate": 44100,
    "rms_energy": 0.04,
    "crest_factor": 11.86,
    "dynamic_spread": 0.06,
    "spectral_centroid": 1476.24,
    "spectral_flatness": 0.01,
    "spectral_bandwidth": 1796.65,
    "estimated_rt60": 2.94
}

# Incontext examples

QUESTION: please design a reverb audio effects for a echo piano sound.
ANSWER: {'reverb': {'band0_gain': 0.0, 'band1_gain': 0.0, 'band2_gain': 0.0, 'band3_gain': 0.0, 'band4_gain': 0.0, 'band5_gain': 0.0, 'band6_gain': 0.0, 'band7_gain': 0.0, 'band8_gain': 0.0, 'band9_gain': 0.0, 'band10_gain': 0.0, 'band11_gain': 0.0, 'band0_decay': 0.1, 'band1_decay': 0.1, 'band2_decay': 0.1, 'band3_decay': 0.1, 'band4_decay': 0.1, 'band5_decay': 0.1, 'band6_decay': 0.1, 'band7_decay': 0.1, 'band8_decay': 0.1, 'band9_decay': 0.1, 'band10_decay': 0.1, 'band11_decay': 0.1, 'mix': 0.8}}

QUESTION: please design a reverb audio effects for a warm piano sound.
ANSWER: {'reverb': {'band0_gain': 0.05, 'band1_gain': 0.1, 'band2_gain': 0.15, 'band3_gain': 0.2, 'band4_gain': 0.25, 'band5_gain': 0.3, 'band6_gain': 0.35, 'band7_gain': 0.4, 'band8_gain': 0.45, 'band9_gain': 0.5, 'band10_gain': 0.55, 'band11_gain': 0.6, 'band0_decay': 1.2, 'band1_decay': 1.8, 'band2_decay': 2.5, 'band3_decay': 3.5, 'band4_decay': 4.5, 'band5_decay': 6.0, 'band6_decay': 7.5, 'band7_decay': 9.5, 'band8_decay': 11.5, 'band9_decay': 14.0, 'band10_decay': 16.5, 'band11_decay': 19.5, 'mix': 0.8}}

QUESTION: please design a reverb audio effects for a distorted guitar sound.
ANSWER: {'reverb': {'band0_gain': 0.05, 'band1_gain': 0.1, 'band2_gain': 0.15, 'band3_gain': 0.2, 'band4_gain': 0.25, 'band5_gain': 0.2, 'band6_gain': 0.15, 'band7_gain': 0.1, 'band8_gain': 0.05, 'band9_gain': 0.02, 'band10_gain': 0.01, 'band11_gain': 0.0, 'band0_decay': 1.0, 'band1_decay': 0.8, 'band2_decay': 0.6, 'band3_decay': 0.4, 'band4_decay': 0.2, 'band5_decay': 0.1, 'band6_decay': 0.05, 'band7_decay': 0.02, 'band8_decay': 0.01, 'band9_decay': 0.005, 'band10_decay': 0.002, 'band11_decay': 0.001, 'mix': 0.8}}

QUESTION: please design a reverb audio effects for a echo guitar sound.
ANSWER: {'reverb': {'band0_gain': 0.0, 'band1_gain': 0.1, 'band2_gain': 0.2, 'band3_gain': 0.3, 'band4_gain': 0.3, 'band5_gain': 0.2, 'band6_gain': 0.1, 'band7_gain': 0.05, 'band8_gain': 0.02, 'band9_gain': 0.01, 'band10_gain': 0.01, 'band11_gain': 0.01, 'band0_decay': 0.1, 'band1_decay': 0.2, 'band2_decay': 0.3, 'band3_decay': 0.4, 'band4_decay': 0.5, 'band5_decay': 0.6, 'band6_decay': 0.7, 'band7_decay': 0.8, 'band8_decay': 0.9, 'band9_decay': 1.0, 'band10_decay': 1.1, 'band11_decay': 1.2, 'mix': 0.7}}

QUESTION: please design a reverb audio effects for a echo drums sound.
ANSWER: {'reverb': {'band0_gain': 0.0, 'band1_gain': 0.0, 'band2_gain': 0.0, 'band3_gain': 0.0, 'band4_gain': 0.0, 'band5_gain': 0.0, 'band6_gain': 0.0, 'band7_gain': 0.0, 'band8_gain': 0.5, 'band9_gain': 1.0, 'band10_gain': 1.0, 'band11_gain': 0.5, 'band0_decay': 0.1, 'band1_decay': 0.1, 'band2_decay': 0.1, 'band3_decay': 0.1, 'band4_decay': 0.1, 'band5_decay': 0.1, 'band6_decay': 0.1, 'band7_decay': 0.1, 'band8_decay': 0.7, 'band9_decay': 0.9, 'band10_decay': 0.9, 'band11_decay': 0.7, 'mix': 0.8}}

QUESTION: please design a reverb audio effects for a church guitar sound.
ANSWER: 
\end{lstlisting}

\newpage

\subsection{Output example of LLM inference}
The LLM generates structured JSON outputs containing numerical parameters for direct application to the audio processing chain.

\begin{lstlisting}[style=docstring]
{
    "reverb": {
        "band0_gain": 0.0,
        "band1_gain": 0.1,
        "band2_gain": 0.2,
        "band3_gain": 0.3,
        "band4_gain": 0.4,
        "band5_gain": 0.5,
        "band6_gain": 0.6,
        "band7_gain": 0.7,
        "band8_gain": 0.8,
        "band9_gain": 0.9,
        "band10_gain": 1.0,
        "band11_gain": 0.9,
        "band0_decay": 3.0,
        "band1_decay": 2.5,
        "band2_decay": 2.0,
        "band3_decay": 1.5,
        "band4_decay": 1.2,
        "band5_decay": 1.0,
        "band6_decay": 0.8,
        "band7_decay": 0.6,
        "band8_decay": 0.4,
        "band9_decay": 0.3,
        "band10_decay": 0.2,
        "band11_decay": 0.1,
        "mix": 0.7
    }
}
\end{lstlisting}

\end{document}